\documentclass[reprint,
nofootinbib,
amsmath,amssymb, aps
]{revtex4-2}

\pdfoutput=1
\usepackage[font=small, labelfont=bf]{caption}
\usepackage{booktabs}
\usepackage{graphicx}
\usepackage{bm}
\usepackage{physics}
\usepackage{mathtools}
\usepackage{mathrsfs}
\usepackage{float}
\usepackage{hyperref}

\usepackage{xcolor}

\newcommand\numeqarrow[1]%
{\stackrel{\scriptscriptstyle(\mkern-1.5mu#1\mkern-1.5mu)}{\rightarrow}}
\newcommand\numeqequal[1]%
{\stackrel{\scriptscriptstyle(\mkern-1.5mu#1\mkern-1.5mu)}{=}}

\begin{document}

\preprint{APS/123-QED}

\title{Functional Renormalisation Group for Brownian Motion I: \\ The Effective Equations of Motion}

\author{Ashley Wilkins}
\email{a.wilkins3@newcastle.ac.uk}
\author{Gerasimos Rigopoulos}%
\email{gerasimos.rigopoulos@newcastle.ac.uk}
\affiliation{School of Mathematics, Statistics and Physics, Newcastle University,\\Newcastle upon Tyne, NE1 7RU, United Kingdom
}%
\author{Enrico Masoero}
\email{enrico.masoero@newcastle.ac.uk}
\affiliation{School of Engineering, Newcastle University,\\Newcastle upon Tyne, NE1 7RU,  United Kingdom}

\date{\today}

\begin{abstract}
\noindent We use the functional Renormalisation Group (fRG) to describe the in and out of equilibrium dynamics of stochastic processes, governed by an overdamped Langevin equation. Exploiting the connection between Langevin dynamics and supersymmetric quantum mechanics in imaginary time, we write down renormalisation flow equations for the effective action, approximated in terms of the Local Potential Approximation and Wavefunction Renormalisation. We derive \textit{effective equations of motion} (EEOM) from the effective action (EA) $\Gamma$ for the average position $\left\langle x\right\rangle$, variance $\langle \left(x-  \langle x \rangle\right)^2\rangle$ and covariance.  The fRG flow equations outlined here provide a concrete way to compute the EA and thus solve the derived EEOM. The obtained effective potential should determine directly the  exact equilibrium statistics, name the position, the variance, as well as all higher order cumulants of the equilibrium Boltzmann distribution. This first paper of a two part series is mostly concerned with setting up the necessary formalism while in part two we will numerically solve the equations derived her and assess their validity both in and out of equilibrium.  
\end{abstract}

\maketitle


\section{\label{sec:Intro}Introduction}
Stochastic processes appear in all kinds of contexts in physics. From the Brownian motion of small particles in a thermal bath \cite{VanKampen2007, gardiner2009stochastic} to scalar fields experiencing quantum fluctuations in the early inflationary universe \cite{Starobinsky1994}, many problems of interest can be described by the overdamped Langevin equation (\ref{eq:langevin}). However, the fluctuations (thermal or effectively thermal) occur very frequently and if one were to attempt to adequately simulate such a process a suitable small timestep size would have to be chosen. This means we only have an immediate understanding of the physics on small timescales. Understanding long-time behaviour and finding the equilibrium properties of the system from its initial out-of-equilibrium state requires following the stochastic process for times much longer than this fundamental timescale. It is natural therefore to ask if a `coarse grained' description in time would be beneficial in tracking the long time behaviour at reduced computational cost. This desire to coarse-grain time and examine physics on different temporal scales lends itself naturally to the tools of the Renormalisation Group (RG). 

The renormalisation group was brought to full force through the work of K. Wilson \cite{Wilson1983} who used it to understand phase transitions and since then the RG has become a widely used technique in modern physics with many applications in both particle physics \cite{Peskin:1995ev} and condensed matter physics \cite{Chaikin1995}. 
The RG is relevant whenever fluctuations significantly influence the state (static or dynamical) of a physical system. Its conceptual framework as applied in condensed matter is perhaps most apt for describing the goal in this work: the RG interpolates between a small lattice size, where the underlying physics is known, to a much larger lattice size by including the effect of fluctuations on all intermediate length scales, providing an \textit{effective} picture that averages over all such fluctuations. In this work we apply this idea to the stochastic dynamics of a Brownian particle. For us the small lattice size corresponds to a small fundamental timescale over which the dynamics is adequately described by the Langevin equation (\ref{eq:langevin}). We seek an effective description, valid over much longer timescales, that captures the aggregate effect of fluctuations. The effective description is embodied in an \emph{effective action} $\Gamma[\chi(t)]$ of the average position $\chi(t) \equiv \langle x(t) \rangle$. In particular, one can use the effective action to compute $n$-point correlation functions of the particle's position $\langle x(t_1) x(t_2)\ldots x(t_n) \rangle $, characterizing the system's statistical properties. To obtain this effective long-time behaviour we will use a version of the RG known as the functional or exact or non-perturbative Renormalisation Group. 

The functional Renormalisation Group (fRG) \cite{Wetterich1993, Morris1994} has been applied successfully to many nonequilibrium problems \cite{Canet2004, Gezzi2007, Jakobs2007, Gasenzer2008, Berges2009, Gasenzer2010, Kloss2011, Canet2011a, Sieberer2014, Mukherjee2015, Pawlowski2015, Corell2019}. However this tends to come with many technical difficulties not present in equilibrium systems. Typically one has to begin with an initial state and calculate correlation functions from it using the Schwinger-Keldysh closed time path, sometimes called the in-in formalism. For an introduction to the fRG as applied to nonequilbirium using this technique see e.g. \cite{Berges2012, Berges2016}. What we will outline in this paper however is a much less technically involved method to computing in and out of equilibrium dynamics in systems which exhibit \textit{classical} stochastic fluctuations. In particular we describe how the fRG can be used to derive \textit{effective equations of motion} that will be solved in part two of this series. We preface this derivation with a review of what is known and bring together disparate elements from the literature into a coherent narrative. The busy reader is therefore directed to sections \ref{sec:acc eom} \& \ref{sec:observables} for our main results.

The fRG offers advantages compared to other formulations of the RG, including its ability to deal with theories with strong couplings and its focus on a single object, the effective action $\Gamma$. These features will allow us to study stochastic motion in non-harmonic potentials with arbitrary shapes that do not offer themselves to be studied straightforwardly via standard perturbative RG methods more commonly employed.\footnote{An example of a non-standard perturbative technique that can deal with non-harmonic potentials (such as the doublewell) is the self-consistent expansion \cite{Remez2018} which can also compute equilibrium correlation functions to good accuracy. We would like to thank Eli Barkai for pointing this out to us.} The effective action can be thought of as an analogue to the statistical free energy and can be derived from the partition function or generating functional via a Legendre transform. Wetterich showed \cite{Wetterich1993} - see also \cite{Morris1994} - how one can define $\Gamma$ at some particular energy or momentum scale $\Lambda$ in the UV (small timestep/high frequency for us) where the theory is known and then create an RG flow that interpolates through all energy (frequency/momentum) scales down to the IR (i.e. increasing timestep size/decreasing frequency). This changing of the effective theory at different energy scales or lattice sizes, the fundamental idea behind the RG, can be formulated in a differential equation known as the Wetterich equation:
\begin{equation}\label{Wetterich eqn}
k\partial_{k}\Gamma_{k} = \dfrac{1}{2}\textbf{Tr}\left[k\partial_{k}R_{k}\left(\Gamma_{k}^{(2)} + R_k\right)^{-1} \right]
\end{equation}
where $\Gamma_{k}$ is the effective action at energy scale (frequency/momentum) $k$, $\textbf{Tr}$ denotes a trace over spatio-temporal points (an integral over spacetime) and a trace over all other relevant indices, $R_{k}$ is an IR regulator that acts as a cut-off for fluctuations below energy scale/frequency $k$, and  $\Gamma_{k}^{(2)}$ is the second functional derivative of $\Gamma_{k}$ -- see \cite{Berges2002} for a review and an entry point to the literature on the subject, \cite{Dupuis2020} for a comprehensive overview of applications as well as e.g. \cite{Gies2012, Delamotte2012} for more elementary introductions. A simple manifestation of this flow equation, stemming from the Boltzmann equilibrium distribution, is presented in Appendix \ref{app:equil-flow}.  In this work we put equation (\ref{Wetterich eqn}) to use for studying the dynamics of particles under the influence of a deterministic force, stemming from an arbitrary potential, and thermal fluctuations. 

We start in Sec. \ref{sec:Brown} by reviewing the connection between Langevin dynamics and Supersymmetric quantum mechanics in imaginary time first shown in \cite{Niel1987} - see e.g. \cite{zinn2002quantum} for a review of this connection. The path integral formulation of section \ref{sec:Brown} then allows us to apply the fRG program directly. We also include a brief summary of how the Langevin equation can be reformulated in terms of a probability distribution function whose evolution is described by the Fokker-Planck equation (\ref{eq: F-P}) and how the latter relates to a Euclidean Schr\"{o}dinger equation.

In Sec. \ref{sec:fRG} we present the flow equations for the effective action utilising a slight modification of the results of \cite{Synatschke2009} for supersymmetric RG flows. As we explain, the flow equation derived from the supersymmetric formulation ensures compatibility with the equilibrium Boltzmann distribution, in contrast to a naive application of the renormalisation group to the Onsager-Machlup form of the generating functional. Recently, the fRG has been applied for smoothing temporal fluctuations in Langevin dynamics in \cite{Duclut2017} without direct use of supersymmetry. As discussed in \cite{Moreau2020}, the physically inspired conditions the authors of \cite{Duclut2017} require of their flow equations are straightforwardly imposed by the Supersymmetric flow. The supersymmetric flow equation itself was first derived in \cite{Synatschke2009} but without making any connection to stochastic dynamics. This connection was made independently in \cite{Canet2011} -- see also \cite{Damgaard1987} -- which however considered a field theory in extended spatial dimensions and smoothing of spatial fluctuations, not temporal fluctuations as we do here. In fact, the authors of \cite{Synatschke2009} obtain a slightly different flow equation when wavefunction renormalisation is included since they do not connect the action functional they study to Brownian motion and the corresponding equilibrium Boltzamnn distribution. This Supersymmetric fRG flow has only been very recently utilized in the context of the stochastic dynamics of early universe inflation \cite{Prokopec2018, Moreau2020, Moreau2020a}. To turn the functional integro-differential equation (\ref{Wetterich eqn}) into a mathematically more tractable form we employ two commonly used approximations for the effective action $\Gamma_k$: the Local Potential Approximation (LPA) as well as the LPA augmented by Wavefunction Renormalisation (WFR). As we discuss, the LPA offers a clear physical interpretation for the effective action: as the effect of fluctuations is progressively taken into account during the flow, the effective potential $V_k(x)$ experienced by the particle  is altered, compared to the bare, fundamental potential $V(x)$. This physical interpretation is often under-emphasised or absent in the literature hence our desire to highlight it here. Wavefunction Renormalisation (WFR) involves a second function $Z_k(x)$ which can be interpreted as a redefinition of position $x \rightarrow Z(x)$. 

In Sec. \ref{sec:acc eom} we demonstrate how the Effective Action $\Gamma$ allows one to naturally compute so-called \textit{effective} equations of motion (EEOM) through variational derivatives in a manner completely analogous to how one computes the classical equations of motion from the classical action $\mathcal{S}$. We do this first for the one point function (or average position $\chi$) and show how its equation of motion simply reduces to an overdamped equation in an effective potential with no noise i.e. purely classical. We then demonstrate how the evolution of the two point function can be obtained by taking two variational derivatives of $\Gamma$ and solving this Green's function equation (\ref{eq:2pointfuncdef}) gives the effective equation of motion for the Variance (\ref{eq:QEOM Variance}) and Covariance (\ref{eq:QEOM Covariance}).  

In Sec. \ref{sec:observables} we discuss the equilibrium limit and demonstrate how the fRG can straightforwardly provide observable quantities that could be measured in simulations. These are in the form of n-point connected correlation functions, or Ursell functions, which can be obtained from $\Gamma_{k = 0}$ in a standard way. In particular, we demonstrate how the effective potential and wave function renoromalization provide the equilibrium average position, variance and the time dependence of the covariance or connected 2-point function, while also discussing how higher order correlation functions like the 3-point and 4-point functions can in principle be obtained for particles in equilibrium.

We conclude in Sec. \ref{sec:Summary} by summarising our analytical results. In Appendix \ref{app:1D=2D} we show that our results also apply for two mutually interacting particles in 2D and 3D. Appendix  \ref{app:equil-flow} derives the analogue of the Wetterich equation (\ref{Wetterich eqn}) for the effective potential corresponding to the equilibrium Boltzmann distribution. In Appendix \ref{app:2ptfuncderiv} we include a detailed derivation for solving (\ref{eq:2pointfuncdef}) in order to obtain the EEOM for the two-point function. 

Part two of this series will explicitly solve the equations derived here numerically and assess their ability to recover statistical correlators both in and out of equilibrium. We find great agreement at moderate to high temperatures (compared to typical barrier heights of the relevant problem) with the low temperature regime being outside the validity of the fRG derivative expansion approach.
\section{\label{sec:Brown} Brownian Motion as SuperSymmetric Quantum Mechanics}
This section pedagogically reviews Brownian motion and its path integral formulation that resembles SuperSymmetric Quantum Mechanics. We exploit this link to derive the flow equations in the next section but the reader familiar with this formulation of Brownian Motion can skip this section. 

Brownian motion for a single particle of mass $m$ moving in a potential $\bar{V}(x)$, coupled to an external heat bath with temperature $T$, can be described by the Langevin equation:
\begin{eqnarray}
m\ddot{x} + \gamma \dot{x} &=& -\partial_{x} \bar{V}(x) + f(t) \label{eq:langevinfull} \\
\left\langle f(t)f(t') \right\rangle &=& 2D\gamma^2 \delta (t-t')\label{eq:f defn}
\end{eqnarray}
where $\gamma$ is a frictional term due to the surrounding fluid, $f(t)$ is a gaussian ``noise" term and $\bar{V}(x)=m V(x)$ is the particle's potential energy. $D = k_bT/\gamma$ is the diffusion constant with equality given so as to match the Boltzmann equilibrium distribution (should it exist). Hereafter, we will be concerned with the overdamped limit:
\begin{eqnarray}
\dot{x} &=& -\varepsilon \partial_{x} V(x) + \eta(t) \label{eq:langevin} \\
\left\langle \eta(t)\eta(t') \right\rangle &=& 2D \delta (t-t')\label{eq:eta defn}
\end{eqnarray}
to which the system settles over a timescale $\varepsilon \equiv m/\gamma$ which we assume to be short.\footnote{These equation for a single particle in 1-D can also be used to describe the radial separation of two particles moving in 2-D or 3-D -- see Appendix \ref{app:1D=2D}.} Note that the overdamped equations are a consistent approximation to the full dynamics as long as $m^2V''/\gamma^2\ll 1$.

We will be examining the impact of changing the temperature, and hence changing the strength of the fluctuating force $\xi$, on the coarse-grained effective theory. Let us therefore introduce a reference temperature $T_0$ and a dimensionless parameter $\Upsilon$ which allows us to dial the temperature around $T_0$.  Writing $D=D_0\Upsilon$, we further define dimensionless variables 
\begin{equation}
x=\sqrt{2D_0\varepsilon} \,\hat{x}\,,\quad t=\varepsilon \, \hat{t} 
\end{equation}
\begin{equation}
V(x)=\frac{2D_0}{\varepsilon} \, \hat{V}(\hat{x})\,,\quad \eta(t)=\sqrt{\frac{2D_0}{\varepsilon}} \, \hat{\eta}(\hat{t})
\end{equation}
in terms of which the dynamical equation becomes
\begin{eqnarray}
\frac{d\hat{x}}{d\hat{t}} &=& -\frac{\partial\hat{V}}{\partial\hat{x}} + \hat{\eta}(\hat{t}) \label{eq:langevindimless}\\
 \langle \hat{\eta}(\hat{t}) \hat{\eta}(\hat{t}')\rangle &=& \Upsilon \delta(\hat{t} - \hat{t}')
\end{eqnarray}   
From here onwards we will be dropping the hats for simplicity of notation but  generally refer to dimensionless quantities unless otherwise stated. 

\subsection{The Brownian Motion Path Integral}
In order to bring the powerful tools of Quantum Field Theory such as the fRG to bear, we will need to reformulate the stochastic differential equation (\ref{eq:langevindimless}) in terms of a \textit{path integral}. In this subsection we will outline one way to obtain this path integral, aiming to link this to Supersymmetric Quantum Mechanics. Our final expression, and the starting point of our subsequent analysis, is the Brownian Motion transition probability (\ref{eq:TransProb}), expressed in terms of an integral over possible histories weighted by the action (\ref{eq: PDF action}), to which the busy reader may progress if uninterested in the details of the derivation. We will be using a condensed functional notation of infinite dimensional functional integrals but all expressions can be considered as limits of large, finite dimensional ordinary integrals. This derivation is based on the path integral reformulation by De Dominicis, Peliti and Janssen \cite{DEDOMINICIS1976, Janssen1976, DeDominicis1978} of the well known Martin-Siggia-Rose approach for stochastic dynamics, first developed in \cite{Martin1973}.  More details on these path integrals, including the corresponding finite discretisation of the stochastic process can be found in \cite{Lau2007}.

The dynamics of the (dimensionless) Langevin equation (\ref{eq:langevindimless}) can be captured in terms of the \textit{Probability Distribution Function} (PDF) $\mathcal{P}(x_{f}\vert x_{i})$ of observing the particle at $x_f$ at time $t = t_f$ given that initially at $t = t_i$ the particle was at $x_i$. By definition this can be expressed as:
\begin{equation}
\mathcal{P}(x_{f}\vert x_{i}) = \left\langle \delta\left( x(t_f)-x_f \right) \right\rangle \label{eq:PDF defn}
\end{equation}
where the expectation value is taken over all possible realisations of the noise $\eta(t)$ and $\delta\left( x(t_f)-x_f \right) $ is the Dirac delta function. Put another way, $x(t_f)$ is the position at $t_f$ for a given noise history $\eta(t)$ and the brackets indicate averaging over all possible noise histories, or stochastic paths, which start at $x_i$ and end up at $x(t_f)=x_f$ at $t_f$. This is precisely a path integral so we can rewrite the PDF using a gaussian measure for noise (\ref{eq:eta defn}) and express the average as
\begin{eqnarray}
\mathcal{P}(x_{f}\vert x_{i}) = \int\mathcal{D}\eta(t)\delta\left( x(t_f)-x_f \right)\text{exp}\left[-\int \text{d}t \, \dfrac{\eta^2(t)}{2\Upsilon}\right]\nonumber \\
\label{eq: PDF first PI}
\end{eqnarray}
where each noise history is weighted by the exponential factor in the above expression.  We now consider the identity (see e.g. \cite{zinn2002quantum}):
\begin{eqnarray}
1 &=& \int dx_f\int_{x_i}^{x_f}\mathcal{D}x(t)\,\delta\left( x(t)-x_\eta(t)\right)\\
& = & \int dx_f\int_{x_i}^{x_f}\mathcal{D}x(t)\,\delta\left( \dot{x} + V_{,x} -\eta (t)\right) \text{det}\textbf{M}\nonumber \\
&=& \int_{x_i}\mathcal{D}x(t)\,\delta\left( \dot{x} + V_{,x} -\eta (t)\right) \text{det}\textbf{M} \label{eq: useful ident}
\end{eqnarray}
where the matrix $\textbf{M}(t,t')$ is:
\begin{eqnarray}
\textbf{M} &\equiv &\dfrac{\delta \eta(t)}{\delta x(t')} = \left(\dfrac{d}{dt} +  V_{,xx}\right) \delta(t-t')\label{eq: M defn}\,.
\end{eqnarray}
This identity expresses the obvious fact that, if the particle starts at some $x_i$ and follows a particular history $x_\eta(t)$ dictated by the Langevin equation without disappearing, it will end up somewhere after time $t_f$. We have used the standard subscript notation to denote derivative with respect to that variable e.g. $V_{,xx} = \partial_{xx}V$.
Note that the second path integral in (\ref{eq: useful ident}) is over all paths starting at $x_i$ at $t_i$ and ending at any $x$ at $t_f$. Inserting our `fat unity' factor (\ref{eq: useful ident}) into (\ref{eq: PDF first PI}) and noting that the delta function there restricts $x(t_f)$ to be $x_f$ we obtain:
\begin{eqnarray}
\mathcal{P}(x_{f}\vert x_{i}) &=& \int\limits_{x(t_i)=x_i}^{x(t_f)=x_f}\mathcal{D}\eta\mathcal{D}x\,\delta\left( \dot{x} + V_{,x} -\eta \right) \text{ det}\textbf{M}\nonumber \\
&& \times \text{ exp}\left[-\int \text{d}t \,\dfrac{\eta^2(t)}{2\Upsilon}\right] \label{eq: PDF 2nd PI}
\end{eqnarray}
where the $\mathcal{D}x(t)$ integral is taken over all paths beginning at $x_i$ and ending at $x_f$. 
We can rewrite the delta function as a functional Fourier transform using a new variable $\tilde{x}$ which is usually called the \textit{response field}:
\begin{eqnarray}
\delta\left( \dot{x} + V_{,x} -\eta \right) = \int \mathcal{D}\tilde{x}\text{ exp}\left[ i\int \text{d}t~\tilde{x}\left( \dot{x} + V_{,x} -\eta \right)\right] \nonumber \\
\label{eq: tilde x PI}
\end{eqnarray}
There are a couple of standard ways we can incorporate $\rm{det} \textbf{M}$ into an exponential. Formally writing  
\begin{equation}
\textbf{M}=\left(\frac{d}{dt}\right)\left(1+\left(\frac{d}{dt}\right)^{-1} V_{,xx}\right)\equiv\left(\frac{d}{dt}\right)\tilde{\textbf{M}}\,,
\end{equation}
where the matrix $\left(\frac{d}{dt}\right)^{-1}(t,t')=\Theta(t-t')$, we see that 
\begin{eqnarray}
\text{det} \textbf{M} &=& \text{det}\left(\frac{d}{dt}\right) \times \text{det} \tilde{\textbf{M}} \propto\text{exp}\left[ \text{Tr}\,\text{log}\left(\tilde{\textbf{M}} \right)\right] \nonumber\\
&\propto& \text{exp}\left[\dfrac{1}{2}\int \text{d}t~V_{,xx}\right]\label{eq:det-Alg}
\end{eqnarray}
where we used the Stratonovich prescription $(\theta (0) = 1/2)$.
Alternatively, and to make the link with SUSY clearer, we can use anticommuting variables\footnote{These suggestively already look like fermionic fields which we will see they are related to in the Supersymmetric picture} $c$ and $\bar{c}$  such that:
\begin{eqnarray}
\text{det } \textbf{M} = \int \mathcal{D}c\mathcal{D}\bar{c}\text{ exp}\left[\int\text{d}t \, \bar{c}\left( \partial_{t} +  V_{,xx} \right)c \right] \label{eq: ccbar PI}
\end{eqnarray}
This Gaussian integral can be done explicitly also leading to (\ref{eq:det-Alg}). However it pays to keep the determinant expressed in this form. Inserting equations (\ref{eq: tilde x PI}) \& (\ref{eq: ccbar PI}) into (\ref{eq: PDF 2nd PI}) we obtain:
\begin{eqnarray}
\mathcal{P}(x_{f}\vert x_{i}) &=& \int\mathcal{D}\eta\mathcal{D}x\mathcal{D}\tilde{x} \mathcal{D}c\mathcal{D}\bar{c} \nonumber \\
&&\text{ exp}\Bigg[\int\text{d}t \Big\{-\dfrac{\eta^2}{2\Upsilon} + i\tilde{x}\left( \dot{x} + V_{,x} -\eta \right) \nonumber \\
&&\quad \quad \quad + \bar{c}\left( \partial_{t} +  V_{,xx} \right)c \Big\} \Bigg]
\end{eqnarray}
We can now trivially perform the gaussian integral over $\eta$ to obtain the path integral in terms of the Brownian Motion (BM) action $\mathcal{S}_{BM}(x,\tilde{x},\bar{c},c)$:
\begin{eqnarray}
\mathcal{P}(x_{f}\vert x_{i}) &=& \int\mathcal{D}x\mathcal{D}\tilde{x} \mathcal{D}c\mathcal{D}\bar{c} \text{ exp}\left[-\mathcal{S}_{BM}(x,\tilde{x},\bar{c},c)\right] \nonumber \label{eq:TransProb} \\
\label{eq: PDF final}\\
\mathcal{S}_{BM}(x,\tilde{x},\bar{c},c) &=& \int \text{d}t\bigg[ \frac{\Upsilon}{2}\tilde{x}^2 - i\tilde{x}(\dot{x}+ V_{,x}) \nonumber \\
& & \quad \quad \quad - \bar{c}\left( \partial_{t} +  V_{,xx} \right)c \bigg] \label{eq: PDF action}
\end{eqnarray} 
Computing this path integral, which henceforth shall be called the \textit{Brownian Path Integral} (BPI), is in general impossible analytically. Instead we will be using the fRG to compute it numerically using the appropriate flow equation. Redefining our fields as :
\begin{eqnarray}
x(t) &\equiv & \sqrt{\Upsilon}\,\varphi(t) \nonumber \\
V(x) &\equiv & {\Upsilon} \, W(\varphi) \nonumber \\
\tilde{x} &\equiv & \dfrac{1}{\sqrt{\Upsilon}}\,(i\dot{\varphi} - F)\nonumber \\
\bar{c}c &\equiv & i\bar{\psi}\psi \nonumber \\ \label{eq:LPAidentifications}
\end{eqnarray}
we obtain 
\begin{equation}
\mathcal{S}_{BM}[\varphi, F, \bar{\psi}, \psi] = \left[W(\varphi_f) -  W(\varphi_i)\right] + \mathcal{S}_{SUSY}
\end{equation}
where
\begin{eqnarray}
\mathcal{S}_{SUSY}[\varphi, F, \bar{\psi}, \psi] = \int dt\bigg[&\dfrac{1}{2}&\dot{\varphi}^2 + \dfrac{1}{2}F^2 + iFW_{,\varphi}(\varphi) \nonumber \\
&-&i \bar{\psi}(\partial_{t} +  W_{,\varphi\varphi}(\varphi))\psi\bigg]  \label{eq:SUSYClass}
\end{eqnarray} 

Action (\ref{eq:SUSYClass}) describes the dynamics of Euclidean, or imaginary time, Supersymmetric Quantum Mechanics where $\psi$ \& $\bar{\psi}$ are the fermionic fields and $\varphi$ \& $F$ are the bosonic fields \cite{Synatschke2009}. We have shown that the same action also describes Brownian motion and the BM action is equivalent to the SUSY QM one up to a factor depending on the initial and final positions $x_i$ \& $x_f$. As the integrand of the BPI (\ref{eq: PDF final}) does not depend on the final or initial states this (now exponential) factor can be simply taken outside the integral. Variation of $\mathcal{S}_{SUSY}$ with respect to $F$ yields its ``equation of motion'' $F=-iW_{,\varphi}$ which when substituted back into $\mathcal{S}_{SUSY}$ yields the ``on mass-shell'' action
\begin{eqnarray}
\mathcal{S}_{OM}[\varphi, \bar{\psi}, \psi] = \int dt\bigg[&\dfrac{1}{2}&\dot{\varphi}^2 + \dfrac{1}{2}W_{,\varphi}{}^2 \nonumber \\
&-&i \bar{\psi}(\partial_{t} +  W_{,\varphi\varphi})\psi\bigg]
\end{eqnarray} 
It is illuminating to express the above action in terms of the original dimensional variables and perform the integration over $\bar{\psi}$ and $\psi$, leading to the alternative form of the term stemming from the determinant:
\begin{eqnarray}
\mathcal{S}_{OM}[x] = \int \frac{dt}{2Dm}\bigg[&\dfrac{1}{2}&m\dot{x}^2 + \dfrac{1}{2}\varepsilon^2 mV_{,x}{}^2 - Dm\varepsilon V_{,xx}\bigg] \nonumber \\\label{eq:dim-action}
\end{eqnarray} 
Note that $2Dm$ has the dimensions of action and therefore plays in the thermal problem a role analogous to $\hbar$ in quantum mechanics - see also section \ref{sec:F-P} in this respect. Unlike $\hbar$ of course, it can be varied by changing the temperature, therefore controlling the strength of fluctuations.

The fRG was first applied to a system governed by the action (\ref{eq:SUSYClass}) by Synatschke et.~al \cite{Synatschke2009} whose approach we adopt in what follows - see also \cite{Canet2011}. As the the relevant manipulations are quite involved, we refer the reader to \cite{Synatschke2009} for technical details regarding the derivation of the flow equations.  

Before moving on to the fRG we outline how the on-mass shell action (\ref{eq:dim-action}) can be obtained from the Fokker-Planck equation which resembles a Euclidean Schr\"{o}dinger equation.
\subsection{\label{sec:F-P}The Fokker-Planck equation}
Instead of working with the Langevin equation directly once can deal directly with the probability distribution of position:
\begin{equation}
P(x,t) = \left\langle \delta (x-x_{\eta})\right\rangle
\end{equation} 
where $x_{\eta}$ is the solution to (\ref{eq:langevin}) for a given noise function $\eta$ (i.e. a specific trajectory). It can be shown that this evolves according to the following PDE:
\begin{equation}
\dfrac{\partial P(x,t)}{\partial t} = \partial_{x}(P(x,t)\partial_x V) + \dfrac{\Upsilon}{2}\partial_{xx}^{2} P(x,t) \label{eq: F-P}
\end{equation}
which is known as the Fokker-Planck (F-P) equation. It is usually more useful however to rescale the PDF like so:
\begin{equation}
P(x,t) = e^{-V/\Upsilon}\tilde{P}(x,t) \label{eq:Ptransform}
\end{equation}
where the time independent solution $P(x) = e^{-V/\Upsilon}$ which corresponds to (the square root of) the equilibrium Boltzmann distribution, is scaled out. This enables the F-P equation to take the form:
\begin{eqnarray}
\dfrac{\Upsilon}{2}\dfrac{\partial\tilde{P}(x,t)}{\partial t} &=& \left(\dfrac{\Upsilon}{2}\right)^2\partial_{xx}^{2}\tilde{P}(x,t)  + \bar{U}\tilde{P}(x,t) \label{eq: rescaled F-P}\\
\bar{U} &\equiv & \dfrac{\Upsilon}{4}\partial_{xx}^{2} V - \dfrac{1}{4}(\partial_{x} V)^2
\end{eqnarray}
which resembles a Euclidean Schr\"{o}dinger equation with $\Upsilon /2$ playing the role of $\hbar$ to control the fluctuations as one might expect. Following standard procedures, we can then write solutions to this Schr\"{o}dinger equation as a path integral. For our F-P equation we can then express the propagator as (restoring the original dimensional variables):
\begin{equation}
\begin{split}
&\left\langle x_{f},t_{f}|x_{i},t_{i}\right\rangle = \mathcal{N}\text{ exp}\left(\dfrac{\varepsilon}{2D}\left[V(x_f)-V(x_i)\right]\right) \\
&\times\int \mathcal{D}x(\tau) \text{ exp}\left( -\int \dfrac{d\tau}{2Dm} \left\{\dfrac{1}{2}m(\partial_{\tau}{x})^2 - \bar{U}(x)\right\}\right) \label{eq:PI F-P}
\end{split}
\end{equation}
The exponential prefactor has come from the fact that we redefined our physical probability P(x,t) in equation (\ref{eq:Ptransform}) and we must rescale back in order to get the physical probability. We therefore recover the ``on mass-shell" path integral (\ref{eq:dim-action}) obtained earlier. Note the importance of treating the determinant (\ref{eq:det-Alg}) correctly in order to obtain the $\partial_{xx}^2V$ term in the Schr\"{o}dinger potential $U$. 

A benefit of working directly with the probability distribution is that it is straightforward to compute statistical moments. For instance the one- and two-point functions at time T using the PDF are given by:
\begin{eqnarray}
\chi (T) &=& \int_{-\infty}^{\infty} x(T) \cdot P(x,T) dx \label{eq: chifromF-P}\\
\textbf{Var} [x(T)] &=& \int_{-\infty}^{\infty} [x(T)-\chi(T)]^2 \cdot P(x,T) dx \nonumber \\
\end{eqnarray}
Higher statistical cumulants can be computed in a similar manner. If one wants to compute how statistical quantities such as average position evolve out of equilibrium then the standard recipe is as follows:
\begin{enumerate}
\item Start with an initial probability distribution for the particle's position at t = 0: $P_0 = P(x,t = 0)$
\item Evolve this probability distribution according to (\ref{eq: rescaled F-P}) to the desired final time T
\item At each time increment compute your desired statistic using e.g. (\ref{eq: chifromF-P}) for average position
\end{enumerate}
\section{\label{sec:fRG}Applying the functional Renormalisation Group}
The formulation of the fRG involves a functional (infinite dimensional) differential equation known as the Wetterich equation \cite{Wetterich1993} that describes the `flow' of the effective action between the microscopic and macroscopic scale. This `flow' is described by a parameter $k$ that ranges from the UV cutoff $\Lambda$ down to the IR regime as $k$ $\rightarrow$ 0. In our Brownian motion scenario, microscopic regime refers to a small timestep and macroscopic to a long timestep. The definition of $\Lambda \sim 1/\Delta t$ is analogous to the Condensed Matter interpretation of the cutoff being inversely proportional to the lattice size, the only difference here being that the Condensed Matter lattice is in space and ours is in time. We will use the fRG ultimately to calculate correlation functions of the particle position. As this derivation uses known techniques and results we refer the busy reader to our basic equations and main results of this section: equation (\ref{eq:dV/dk}) for the Local Potential Approximation to the RG flow and when we also include Wavefunction Renormalisation they are (\ref{eq:dV/dktilde2}) and (\ref{eq:dzetax/dktilde}). Before starting this derivation, we briefly recall how to generate correlation functions in the standard Field Theory way.

In Euclidean Quantum Field Theory correlation functions can be evaluated with the help of \textit{generating functionals}. The most straightforward of these is the partition functional $\mathcal{Z}(J)$ which depends on a source term $J(x)$ (in analogy with a magnetic field source term $M(x)$ in spin systems). For example, the two point correlation function is:
\begin{eqnarray}
\left\langle x(t_{1})x(t_{2})\right\rangle &\equiv&  \frac{\int\mathcal{D}x\,\, x(t_1)x(t_2)\,\,\text{ exp}\left[-\mathcal{S}[x] \right]}{\int\mathcal{D}x \text{ exp}\left[-\mathcal{S}[x] \right]}\\
&=&\dfrac{1}{\mathcal{Z}(0)} \dfrac{\delta^2\mathcal{Z}(J)}{\delta J(t_{2})\delta J(t_{1})}\bigg\rvert_{J = 0}
\end{eqnarray}
where
\begin{equation}
\mathcal{Z}(J) = \int\mathcal{D}x \text{ exp}\left[-\mathcal{S}[x]  + \int_t J x\right]\,,
\end{equation}
$\mathcal{S}[x]$ is the action and $\int_t=\int dt$. We have assumed in the above that $\mathcal{S}$ depends only on a single variable $x(t)$ for notational brevity but the above formulae are modified straightforwardly for any number of variables $x_i(t)$ which can be coupled to corresponding sources $J_i(t)$. For example $x_i(t)\equiv \left( x(t),\tilde{x}(t),\bar{c}(t),c(t)\right)$ in (\ref{eq: PDF final}).

We can store the information encoded in $\mathcal{Z}(J)$ better in the object $\mathcal{W}[J]$:
\begin{equation}
\mathcal{W}[J] \equiv \text{ln}\left(\mathcal{Z}(J)\right)
\end{equation}
which is the generator of connected correlation functions (or Ursell functions):
\begin{equation}
\left\langle x(t_1)...x(t_n)\right\rangle_{C} = \dfrac{\delta^n \mathcal{W}[J]}{\delta J(t_1)...\delta J(t_n)}
\end{equation}
So for instance the connected 2-point function (more commonly known as covariance) $G(t_1,t_2)$ is:
\begin{eqnarray}
G(t_1,t_2) \equiv \left\langle x(t_1)x(t_2) \right\rangle_{C} &=& \left\langle x(t_1)x(t_2) \right\rangle - \left\langle x(t_1) \right\rangle\left\langle x(t_2) \right\rangle \nonumber \\
&=& \dfrac{\delta^2 \mathcal{W}[J]}{\delta J(t_1)\delta J(t_2)}
\end{eqnarray}
Computing $\mathcal{W}[J]$ directly however is very difficult. Standard approaches involve perturbative expansions leading to the well known diagrammatic Feynman rules. In this work we will calculate a related object, the \textit{effective action} $\Gamma[\chi]$ given by the Legendre transform of $\mathcal{W}[J]$
\begin{eqnarray}\label{effective action}
\Gamma [\chi] = \int_t J\chi - \mathcal{W}[J] 
\end{eqnarray} 
where the field $\chi$ corresponds to the expectation value of $x$ in the presence of the source field $J$, satisfying
\begin{equation}
\chi = \dfrac{\delta \mathcal{W}[J]}{\delta J} = \left\langle x \right\rangle_{J}
\end{equation}

The fRG formulation adds a regulating term to the action in our definition of the generating functional:
\begin{equation}
\mathcal{Z}_k(J) = \int\mathcal{D}x \text{ exp}\left[-\mathcal{S}[x] - \Delta\mathcal{S}_{k}[x] + \int_t J x\right] 
\end{equation}
where the regulating term $\Delta\mathcal{S}_{k}[x]$ is quadratic in $x$:
\begin{equation}
\Delta\mathcal{S}_{k}[x] = \dfrac{1}{2}\int_{t,t'}x(t)R_{k}(t,t')x(t') 
\end{equation}
Crucially $R_{k}$ is an IR regulator that depends on a Renormalisation scale $k$ and the momentum/frequency $p$ of the modes. The precise form of $R_{k}$ is not crucially important and it is chosen in order to optimize calculations but it should suppress  IR modes and vanish as $k \rightarrow 0$, $\lim\limits_{k\rightarrow 0} R_k=0$, ensuring that the full effective action (\ref{effective action}) is recovered in this limit. By defining the mean position as before $\chi(t) \equiv \left< x(t)\right>$ we can construct the Regulated Effective Action:
\begin{equation}
\Gamma_{k}[\chi] = \int_t J \chi - \mathcal{W}_{k}[J]-\Delta\mathcal{S}_{k}[\chi] 
\end{equation}
where $\mathcal{W}_{k}[J]= \text{ln}\left(\mathcal{Z}_{k}\right)$ is analogous to the non-regulated case. 

From the Regulated Effective Action one can obtain obtain the Wetterich equation \cite{Wetterich1993, Morris1994}:
\begin{equation}\label{Wetterich functional}
\partial_{k}\Gamma_{k}[\chi] = \dfrac{1}{2}\int_{t,t'}\partial_{k}R_{k}(t,t')\left[R_{k} + \Gamma_{k}^{(2)}\right]^{-1} 
\end{equation}
which is a functional equation determining how $\Gamma_k$ changes as $k \rightarrow 0$. It interpolates $\Gamma_{k}$ from the microscopic scale ($k = \Lambda$), where $\Gamma_{\Lambda} =\mathcal{S}$, down to the IR regime ($k  = 0$) where the full effective action $\Gamma[\chi] = \Gamma_{k=0}[\chi]$, encoding the effect of all fluctuations, is obtained. A simplified derivation for one degree of freedom at equilibrium, which however captures all the relevant manipulations, can be found in Appendix \ref{app:equil-flow}. 

As demonstrated in the previous section, our Brownian motion problem is actually SUSY QM. We can therefore apply the fRG technology and incorporate the effect of thermal fluctuations by following the flow of the effective action $\Gamma_k$ via the Wetterich equation. Synatschke et. al have analysed a system with action $\mathcal{S}_{SUSY}$ in light of its underlying symmetries in \cite{Synatschke2009}. We adopt their results here. They find that from a supersymmetric perspective, the appropriate regulating term takes the form    
\begin{eqnarray}
\Delta\mathcal{S}_k\! &=&\! \int_{\tau\tau'} \hspace{-0.2cm} r_2(k,\Delta\tau)  \left[-\dot{\phi}(\tau)\dot{\phi}(\tau')+F(\tau)F(\tau')-i\bar{\psi}(\tau)\dot{\psi}(\tau')\right]\nonumber\\
&+&2ir_1(k,\Delta\tau) \left[\phi(\tau)F(\tau')-\bar{\psi}(\tau)\psi(\tau'))\right]
\end{eqnarray}
where $\Delta\tau \equiv \tau-\tau'$. Such a form was also suggested in \cite{Duclut2017}, however we will see that compatibility with the Boltzmann distribution suggests setting $r_{2} \rightarrow 0$. The flow equations of \cite{Synatschke2009} are discussed below. 

\subsection{Local Potential Approximation}

In practice, calculating $\Gamma_k$ exactly is usually impossible and we must consider a truncation to make the functional equation (\ref{Wetterich functional}) tractable. The most common approximation is the so-called derivative expansion. The Local Potential Approximation (LPA), the leading order in the derivative expansion, is the assumption that the only part of the effective action that depends on our momentum scale $k$ is the superpotential $W$.  The effective action then takes the form:
\begin{eqnarray}
\Gamma_{k}[\phi,F,\bar{\psi},\psi] = \int d\tau\bigg[&\dfrac{1}{2}&\dot{\phi}^2 + \dfrac{1}{2}F^2 + iFW_{k, \phi}(\phi) \nonumber \\
&-& i\bar{\psi}\left(\partial_t + W_{k, \phi\phi} \right)\psi \bigg] \label{eq:SUSYEffective}
\end{eqnarray}
such that $\Gamma_{k = \Lambda} = \mathcal{S}_{SUSY}$ under the condition $W_{k = \Lambda}(\phi) = W(\phi)$ with $\phi \equiv \left\langle \varphi\right\rangle$ being the mean field. In this approximation the only thing changing with $k$ directly, progressively incorporating the effect of fluctuations on different timescales, is $W_k$. This means we only have one flow equation to solve which turns out to be \cite{Synatschke2009}:
\begin{equation}
\partial_{k}W_{k}(\phi) = \int_{-\infty}^{\infty}\dfrac{dp}{4\pi}\dfrac{(1+r_2)\partial_{k}r_{1}- \partial_{k}r_{2}~(r_{1} + \partial^2_{\phi}W_{k}(\phi))}{p^2+(r_{1} + \partial^2_{\phi}W_{k}(\phi))^2} \nonumber \\
\end{equation}
We notice that if we set $r_2 = 0$ and choose a local-in-time $r_1(k, \delta\tau)=k\delta(\tau-\tau')$ the so-called Callan-Symanzik regulator then this choice\footnote{Physically speaking the final results should be independent of the regulator chosen. This is a subtlety we will not address in this work as it was shown in \cite{Synatschke2009} that even for other choice of regulators the difference in the final results was negligible, at least for the LPA.} effectively adds a quadratic term to the potential $W \rightarrow W + k\phi^2$ and leads to a relatively simple flow equation: 
\begin{equation}
\partial_{k}W_{k}(\phi) = \dfrac{1}{4}\cdot\dfrac{1}{k+ \partial^2_{\phi}W_{k}(\phi)} \,.
\label{eq:dW/dk}
\end{equation}
In terms of the physical variables we have 
\begin{equation}\label{eq:dV/dk}
\partial_{k}V_{k}(\chi) = \dfrac{\Upsilon}{4}\cdot\dfrac{1}{k+ \partial^2_{\chi}V_{k}(\chi)} \,,
\end{equation}
which shows explicitly the effect of dialling the temperature $\Upsilon$: the higher the temperature the faster the flow as a result of stronger thermal fluctuations. Equation (\ref{eq:dV/dk}) can be discretised in the $\chi$ direction and become a set of coupled ODEs that can be solved in the $k$ direction in order to obtain a numerical solution.  

It is important to note that equation (\ref{eq:dV/dk}) is identical to the flow of the effective potential that corresponds to the equilibrium Boltzmann distribution, see \cite{Guilleux2015} and Appendix C with $R \rightarrow k$. We  therefore see that the form of $\mathcal{S}_{SUSY}$ and deriving flow equations in a framework which respects its symmetries is crucial for establishing consistency with the equilibrium Boltzmann distribution. If one started  directly from the Onsager-Machlup functional (\ref{eq:dim-action}) and naively treated it as an $N=1$ Euclidean scalar theory in one-dimension with the combination $U = \frac{1}{2}\left(V_{,x}\right)^2 - \frac{\Upsilon}{2}V_{,xx}$ as the scalar potential to be evolved along the RG flow, one would have obtained a different flow equation  
\begin{eqnarray}
\partial_{k}U_{k}(\phi) = \dfrac{1}{2}\int_{-\infty}^{\infty}\dfrac{dp}{2\pi}\dfrac{\partial_{k}R_{k}}{p^2+ R_{k} + \partial^2_{\phi}U_{k}(\phi)}\,.
\end{eqnarray}
The corresponding Callan-Symanzik regulator would be  $R_k=k^2$, giving
\begin{equation}
\partial_{k}U_{k}(\phi) = \frac{1}{2} \frac{k}{\sqrt{k^2+\partial^2_{\phi}U_{k}(\phi)}}\,.
\end{equation}
It is unclear how or if the end-of-the-flow potential $U_{k=0}$ from this equation relates to the physical potential $V_{k=0}$ and the flow appears a-priori incompatible with the Boltzmann distribution.

\subsection{Wave Function Renormalisation}

In the previous subsection we assumed that the effective action $\Gamma_k$ only depends on the renormalisation scale through the form of the potential. We now allow for the field $\varphi$ itself to be renormalised which results in a scaling of the kinetic term. The new effective action in the SUSY formalism is \cite{Synatschke2009}:
\begin{eqnarray}
\Gamma_{k}[\phi,\bar{\psi},\psi] = \int dt~\dfrac{1}{2}Z_{,\phi}^{2}\dot{\phi}^{2} + \dfrac{1}{2}\left(\dfrac{W_{,\phi}}{Z_{,\phi}}\right)^2 \nonumber \\
- i\bar{\psi}\left(Z_{,\phi}^2 \partial_{t} + Z_{,\phi} Z_{,\phi\phi}\dot{\phi} - Z_{,\phi\phi} \dfrac{W_{,\phi}}{Z_{,\phi}} + W_{,\phi\phi}\right)\psi
\end{eqnarray}
where we have suppressed the explicit dependence on $k$ of $W$ \& $Z$ to avoid overly cluttered notation. From now on we will in general drop this explicit dependence on $k$ for $W$, $V$, $Z$ \& $\zeta$, defined below, only restoring it when we are directly comparing it to the original cutoff value. We introduce an additional identification in addition to (\ref{eq:LPAidentifications}):
\begin{eqnarray}
\zeta(x) = \sqrt{\Upsilon} Z(\phi) \Rightarrow \zeta_{,x} = Z_{,\phi} \\
\bar{c}c = -i\zeta_{,x}\bar{\psi}\psi
\end{eqnarray}
such that the (on-shell) effective action for Brownian motion is now written as:
\begin{eqnarray}
\Gamma_{k}[\chi,\bar{c},c] = \int dt~\dfrac{1}{2\Upsilon}\zeta_{,\chi}^{2}\dot{\chi}^{2} + \dfrac{1}{2\Upsilon}\left(\dfrac{V_{,\chi}}{\zeta_{,\chi}}\right)^2 \nonumber \\
- \bar{c}\left(\zeta_{,\chi}^2 \partial_{t} + \zeta_{,\chi} \zeta_{,\chi\chi}\dot{\chi} - \zeta_{,\chi\chi} \dfrac{V_{,\chi}}{\zeta_{,\chi}} + \cdot V_{,\chi\chi}\right)c
\end{eqnarray}
The regulator term becomes more complicated for this action and we do not reproduce it here, see \cite{Synatschke2009} for details of this. Following their approach one arrives at the LPA + WFR flow equations: 
\begin{eqnarray}\label{eq:dV/dktilde2}
\partial_{{k}}V_{{k}}(\chi) &=& \dfrac{\Upsilon}{4}\cdot\dfrac{1}{{k}+ \partial^2_{\chi \chi}V_{{k}}(\chi)} \\
\partial_{{k}}\zeta_{,\chi} &=& \dfrac{\Upsilon}{4}\cdot\dfrac{\mathcal{P}}{\zeta_{,\chi}\cdot\mathcal{D}^2 } \label{eq:dzetax/dktilde}\\
\mathcal{D} &\equiv & V_{,\chi\chi} + k\,\zeta_{,\chi}^{2} \\
\mathcal{P} &\equiv & \dfrac{4\zeta_{,\chi\chi} V_{,\chi\chi\chi}}{\mathcal{D}} - \left( \zeta_{,\chi\chi}\zeta_{\chi}\right)_{,\chi} - \dfrac{3\zeta_{,\chi}^{2}V_{,\chi\chi\chi}^{2}}{4\mathcal{D}^2}
\end{eqnarray}
which now consist of the previous LPA equation for the effective potential (\ref{eq:dV/dk}) as expected, augmented by one more flow equation for the wavefunction renormalisation $\zeta_{,\chi}$. 

As before we will integrate the LPA equation (\ref{eq:dV/dk}) by discretising along the $\chi$ direction and solving the resulting set of coupled ODEs in $k$. Once the effective potential $V_k(\chi)$ has been obtained the second PDE can be solved for $\zeta_{,\chi}$ in a similar way. It is worth pointing out here that our approach differs slightly from \cite{Synatschke2009} in that the effective potential obeys the same equation as in the LPA approximation even with the inclusion of WFR.\footnote{For the WFR approximation the authors of \cite{Synatschke2009} use a spectrally adjusted regulator which is evaluated on a background field $\bar{\phi}$. They make the simple choice of identifying this background field with the fluctuation field (i.e. $\bar{\phi} = \phi$). This approach however modifies the flow of $V_k$ -- i.e. equation (\ref{eq:dV/dktilde2}) differs from the LPA version (\ref{eq:dV/dk}) -- which means the flow no longer correctly approaches the Boltzmann equilibrium distribution's effective potential and leads to deviations from the correct equilibrium position and variance. The only choice of $\bar{\phi}$ that prevents this from happening is one where $\mathcal{Z}_{k}'(\bar{\phi}) = 1$ for all $k$ which is what we have done here.} This is because the equilibrium state is described exactly by the LPA equation \cite{Guilleux2015,Guilleux2017, Moreau2020a}, as we mentioned above and explicitly recall in Appendix \ref{app:equil-flow}. The LPA flow equation was first solved in \cite{Synatschke2009, Guilleux2015, Guilleux2017}, while more recently WFR was included for a double well potential in \cite{Moreau2020a}.      

\section{\label{sec:acc eom}The effective equations of motion}

A standard formulation of classical mechanics involves the principle of least action. If one considers the classical action $\mathcal{S}$:
\begin{equation}
\mathcal{S} = \int dt~ L(x,\dot{x})
\end{equation}
where $L(x,\dot{x})$ is the Lagrangian, then one can obtain the equations of motion by taking the variational derivative and setting it equal to zero:
\begin{equation}
\dfrac{\delta \mathcal{S}}{\delta x} = 0 \label{eq:delta S = 0}
\end{equation}
The Effective Action (EA) $\Gamma$ is so named because its definition makes it look like a standard classical action once fluctuations have been integrated out:
\begin{equation}
e^{-\Gamma} = \int\mathcal{D}x~e^{-\mathcal{S}}
\end{equation}
It is then natural to ask whether we can extend the variational principle used to obtain the classical equations of motion from S to obtain \textit{effective equations of motion} from $\Gamma$. As the fRG has $\Gamma$ as its central object it is ideally placed to calculate these effective equations of motion. This is what we will demonstrate in the rest of this section.
\subsection{The EEOM for the one point function}
In a similar way to how the classical action $\mathcal{S}(x)$ can yield the classical equations of motion through variational derivatives, so too does $\Gamma [\chi]$ yield the \textit{effective equation of motion} for the one point function (or average position) $\chi$:
\begin{equation}
\dfrac{\delta\Gamma}{\delta \chi(t)} = 0 \label{eq:gen QEOM}
\end{equation}
Here we have assumed there are no external sources\footnote{N.B. this is not the same as assuming that the noise term (\ref{eq:eta defn}) is zero as this is true for $\Gamma$ by definition} (J = 0).  \\
Under the LPA equation (\ref{eq:gen QEOM}) is:
\begin{equation}
\dfrac{\delta\Gamma_{k = 0}}{\delta \chi(t)} = \ddot{\chi} -  \partial_\chi V_{k=0}(\chi)\,\partial_{\chi}^2V_{k=0}(\chi) = 0
\end{equation} 
where the final equality comes by assuming that source terms have been set to zero (i.e $J(t) = 0$). The WFR version of (\ref{eq:gen QEOM}) reads:
\begin{equation}
\left(\zeta_{,\chi}\dot{\chi}\right)\dot{} - \frac{\partial_\chi V_{k=0}}{\zeta_{,\chi}^2}\left(\partial_{\chi\chi}^2V_{k=0}-\frac{\zeta_{,\chi\chi}}{\zeta_{,\chi}}\partial_\chi V_{k=0}\right) = 0
\end{equation} 
where $\partial_\chi\zeta$ and $\partial^2_\chi\zeta$ are also evaluated at $k=0$.
Both of these second order differential equations can actually be reduced to a first order differential equation like so: 
\begin{eqnarray}
\dot{\chi} = -\tilde{V}_{,\chi}(\chi) \label{eq:QEOM}
\end{eqnarray} 
where we have introduced the \textit{effective dynamical potential} $\tilde{V}$ defined by 
\begin{eqnarray}
\tilde{V}_{\chi}(\chi) \equiv 
\begin{cases}
V_{\chi}(k = 0, \chi), & \text{ for LPA}  \\[10pt]
\dfrac{V_{\chi}(k = 0, \chi)}{\zeta_{\chi}^{2}(k = 0, \chi)}, & \text{ for WFR} 
\end{cases}\label{eq: Vtilde}
\end{eqnarray}
Here we can clearly see that for LPA the \textit{effective} and \textit{effective dynamical} potentials are equivalent whereas WFR receives an additional factor. 

Equation (\ref{eq:QEOM}) tells us that the equation of motion for the average position $\chi$ is an extremely simple first order differential equation that appears like a Langevin equation with no noise. This means that once you have obtained the \textit{effective dynamical} potential you can compute the evolution of the average position $\chi$ trivially from any starting position. We will demonstrate this in part 2 of this series.

\subsection{The EEOM for the two point function}
The connected 2-point  function $G(t,t')=\left\langle x(t)x(t')\right\rangle_{C}=\delta^2\mathcal{W}/\delta J(t_1)\delta J(t_2)$ and the second functional derivative of the effective action $\Gamma_{k=0}$ are inverse to each other
\begin{equation}
\int d\tau~\dfrac{\delta^2\Gamma_{k=0}}{\delta \chi(t)\delta \chi(\tau)}\dfrac{\delta^2\mathcal{W}_{k=0}}{\delta J(\tau)\delta J(t')} = \delta(t-t')\label{eq:2ptW and gamma}
\end{equation} 
Concretely this means that the connected 2-point function $G(t,t')$ satisfies the following equation:
\begin{eqnarray}
\left(\dfrac{d^2}{dt^2} -  \mathcal{U}(\chi) \right)G(t,t') &=& -2\Delta\delta(t-t')  \label{eq:2pointfuncdef} 
\end{eqnarray}
where $\mathcal{U}(\chi)$ is:
\begin{eqnarray}
\mathcal{U}(\chi)= 
\begin{cases}
    V_{,\chi\chi}^{2} + V_{,\chi}V_{,\chi\chi\chi}, & \text{for LPA}\\[10pt]
    \dfrac{V_{,\chi\chi}^{2}}{\zeta_{,\chi}^{4}}+ \dfrac{V_{,\chi}V_{,\chi\chi\chi}}{\zeta_{,\chi}^{4}}- \dfrac{V_{,\chi}^{2}\zeta_{,\chi\chi\chi}}{\zeta_{,\chi}^{5}}  \\
 -\dfrac{5V_{,\chi}V_{,\chi\chi}\zeta_{,\chi\chi}}{\zeta_{,\chi}^{5}}    + \dfrac{5V_{,\chi}^{2}\zeta_{,\chi\chi}^{2}}{\zeta_{,\chi}^{6}} , & \text{for WFR}
  \end{cases} \label{eq:U = }
\end{eqnarray}
and
\begin{eqnarray}
\Delta\equiv
\begin{cases}
\dfrac{\Upsilon}{2}, & \text{for LPA}\\[10pt]
\dfrac{\Upsilon}{2\zeta_{,\chi}^{2}}, & \text{for WFR}
\end{cases}
\label{eq:Deltadef}
\end{eqnarray}
The derivation of the full solution to (\ref{eq:2pointfuncdef}) can be found in Appendix~\ref{app:2ptfuncderiv} but here we just highlight the two main results:\\
The EEOM for the \textit{Variance} $t'\rightarrow t$:
\begin{eqnarray}
\textbf{Var}(x) \equiv G(t,t) &=& \dfrac{\Upsilon}{2\lambda P(t)} \tilde{Y}_1(t)\tilde{Y}_2(t)  \nonumber \\
&& +~ \dfrac{P(0)}{P(t)}\left[G_{00} - \dfrac{\Upsilon}{2\lambda P(0)}\right]\tilde{Y}_{2}^{2}(t)\nonumber \\
 \label{eq:QEOM Variance}
\end{eqnarray}
and the EEOM for the \textit{Covariance} $t'\rightarrow 0$, $t > 0$:
\begin{eqnarray}
\textbf{Cov}(x(0)x(t)) \equiv G(t,0) = G_{00}\tilde{Y}_2(t) \label{eq:QEOM Covariance}
\end{eqnarray}
where $\tilde{Y}_i(t) \equiv Y_i(t)/Y_i(0)$ are the `normalised' solutions to the homogeneous equation (\ref{eq:appendhomo2ptfunc}) which can be obtained numerically. $P(t) = 1$ or $\zeta_{\chi}(\chi(t))$ for LPA and WFR respectively and $\lambda$ is defined by (\ref{eq:lambda}). $G_{00} = G(0,0)$ is the initial variance at $t = 0$

\section{\label{sec:observables}The equilibrium limit}
While the Effective Equations of Motion derived in Section \ref{sec:acc eom} are valid for non-equilibrium evolution it is important to ensure that they converge to the correct equilibrium limit. At equilibrium the equations are greatly simplified resulting in the equilibrium position $\chi_{eq}$ and variance $\textbf{Var}_{eq}(x)$ becoming static quantities as expected. We will also show how the covariance at equilibrium is given by an exponential decay with exponent predicted by the fRG.


\subsection{Equilibrium 1-point function}
\label{sec:1-point function}
At equilibrium the average position of the particle should not change, this means that $\dot{\chi} = 0$. It naturally follows from this condition and the EEOM for $\chi$ (\ref{eq:QEOM}) that equilibrium is defined for both LPA \& WFR by the condition 
\begin{eqnarray}
\partial_\chi V_{k = 0}(\chi_{eq}) = 0 \label{eq:equilbirum position}
\end{eqnarray}  
As the potential $V_{k=0}(\chi)$ should be convex (by definition of $\Gamma$) equation (\ref{eq:equilbirum position}) tells us that $\chi_{eq}$ corresponds to the minimum of  $V_{k=0}(\chi)$. Or more concretely: 
\begin{equation}
\lim\limits_{t \to \infty} \left\langle x(t) \right\rangle =  x \text{ that minimises } V_{k=0}(x)
\end{equation}
The equilibrium position is obviously the same for both LPA and WFR as they both lead to the same effective potential. As the equilibrium position is straightforwardly computed from the Boltzmann distribution verifying that the minimum of the effective potential matches the predicted equilibrium position is a good first test that the procedure we have outlined here is valid.

\subsection{Equilibrium 2-point function}
If we now take the equilibrium limit $\chi \rightarrow \chi_{eq}$ of the full EEOM for the 2-point function (\ref{eq:2pointfuncdef}) we find that it simplifies to:
\begin{eqnarray}
\left(\dfrac{d^2}{dt^2} -  \lambda^2\right)G_{eq}(t_1,t_2) =  -2\Delta \vert\delta(t_2-t_1)  \label{eq:2pointfuncequi} 
\end{eqnarray}
where
\begin{eqnarray}\label{eq:lambdadef}
\lambda^2 \equiv  
\begin{cases}
V_{,\chi\chi}^{2}\vert, & \text{for LPA}\\[10pt]
\dfrac{V_{,\chi\chi}^{2}\vert}{\zeta_{,\chi}^{4}\vert}, & \text{for WFR}
\end{cases} 
\end{eqnarray}
and $\Delta$ is defined as in (\ref{eq:Deltadef}). The notation $\vert$ means we have evaluated the function at $k = 0$ and at equilibrium $\chi = \chi_{eq}$.

The appropriate solution to (\ref{eq:2pointfuncequi}) providing the connected correlation function at equilibrium is  
\begin{eqnarray}
G_{eq}(t_1,t_2) = \textbf{Cov}_{eq}(x(t_1)x(t_2))  &=& \dfrac{\Upsilon}{2V_{,\chi\chi}\vert} e^{-\lambda|t_1-t_2|}\nonumber \\ \label{eq:2-pointfunc} \\
\Rightarrow G_{eq}(t,t) = \textbf{Var}_{eq}(x) &=& \dfrac{\Upsilon}{2V_{,\chi\chi}\vert}
\label{eq:equal2pt}
\end{eqnarray}
As the equilibrium variance is also easily computed from the Boltzmann distribution, equation (\ref{eq:equal2pt}) gives us a second test to verify that the effective potential and by extension fRG recipe we have outlined has physical significance. 
In the LPA approximation the variance and the decay rate of the autocorrelation function are both directly given by the curvature of the effective potential at its minimum. The inclusion of WFR however alters the decay rate without changing the equilibrium variance. This is as it should since the latter is fixed by the equilibrium Boltzmann distribution. As we will see in paper two, where explicit results for various potentials are given, WFR improves the decay rate which is indeed not exactly determined by the effective potential's curvature. 

\subsection{Equilibrium connected 3- \& 4-point functions}
While the EEOM for higher point functions become very complicated out of equilibrium they are much simpler in the equilibrium limit. The connected and 1PI correlation functions can be calculated in a standard way from $\Gamma$(see for e.g. pg 381-382 of \cite{Peskin:1995ev}). Assuming that $t_4 \geq t_3 \geq t_2 \geq t_1$ then the connected 3-point function in equilibrium for example is:
\begin{eqnarray}
\left\langle x(t_1)x(t_2)x(t_3) \right\rangle_{C}  &=& \left\langle x(t_1)^3 \right\rangle_{C} e^{-\lambda (2t_3 -t_2-t_1)} \nonumber \\
\label{3-pointfunc}\\
\left\langle x(t_1)^3 \right\rangle_{C} & = & -\left\langle x(t_1)^2 \right\rangle_{C}^{3}\dfrac{\mathcal{V}_{,\chi\chi\chi}\vert}{3\lambda}
\end{eqnarray}
Where we have introduced the notion of the \textit{bosonic potential} $\mathcal{V}$:
\begin{eqnarray}
\mathcal{V}(k,\chi) \equiv  
\begin{cases}
V_{,\chi}^{2}(k,\chi), & \text{for LPA}\\[10pt]
\dfrac{V_{,\chi}^{2}(k,\chi)}{\zeta_{,\chi}^{2}(k,\chi)}, & \text{for WFR}
\end{cases}
\label{eq:bosondef}
\end{eqnarray}
Similarly the connected 4-point function at equilibrium is:
\begin{eqnarray}
&&\left\langle x(t_1)x(t_2)x(t_3)x(t_4) \right\rangle_{C} = \left\langle x(t_1)^4 \right\rangle_{C}e^{-\lambda (3t_4 -t_3 -t_2-t_1)} \nonumber \\
\label{4-pointfunc}\\
&&\left\langle x(t_1)^4 \right\rangle_{C} =  \dfrac{\left\langle x(t_1)^2 \right\rangle_{C}^{4}}{4\lambda}\left(\mathcal{V}_{,\chi\chi\chi}^{2}\vert\dfrac{\left\langle x(t_1)^2 \right\rangle_{C}}{\lambda} - \mathcal{V}_{,\chi\chi\chi\chi}\vert\right) \nonumber \\
\end{eqnarray}
From these connected correlation functions one should be able to calculate the skewness and kurtosis of the equilibrium distribution in a similar manner to what we did for $\chi_{eq}$ and \textbf{Var}(x). We leave this calculation and comparison to the Boltzmann distribution for future work.

\section{\label{sec:Summary} Summary}

We have demonstrated how Brownian motion can be formally described by a  path integral involving a Euclidean Supersymmetric action and how an effective average action functional $\Gamma[\chi]$ of the average position $\chi$, incorporating the effects of the fluctuating force and encoding all statistical properties of the process, can be calculated using functional Renormalisation Group (fRG) methods. The fRG flow equations were written down for the first two orders of the widely used derivative expansion of the effective action, referred to as the Local Potential Approximation (LPA) and Wavefunction Renormalisation (WFR). We used a particular type of regulator, the frequency independent Callan-Symanzik regulator, for which the flow equations take on a relatively simple form. We further recalled that obtaining flow equations within the supersymmetric framework is crucial for ensuring compatibility with the Boltzmann equilibrium distribution, something that is not a priori guaranteed if one starts with the Onsager-Machlup form of the action (\ref{eq:dim-action}) and and considers it a Euclidean N=1 scalar theory in one dimension with the ``Schr\"{o}dinger" potential $U = 1/2\,(V')^2 - \Upsilon/2 \, V''$.

We have shown how the Effective Action (EA) $\Gamma$ allows one to derive \textit{effective equations of motion} (EEOM) in an analogous manner to the classical equations of motion by taking variational derivatives. The EEOM we computed tell us how the average position $\chi$ and variance $\textbf{Var}(x)$ evolve out of equilibrium. In principle the EEOM for higher statistical moments such as skewness and kurtosis could also be derived by taking further variational derivatives. For Brownian motion described by an overdamped Langevin equation this EA can be computed using functional Renormalisation Group (fRG) techniques under the widely used approximations, the LPA and WFR. We have also outlined how the more standard method of evolving the Fokker-Planck (F-P) diffusion equation can also compute $\chi$ and $\textbf{Var}(x)$. 

We also explored the equilibrium limit further emphasising the physical significance of certain aspects of the effective potential $V_{k=0}$. Namely how the minimum of $V_{k=0}$ corresponds to the equilibrium position and its second derivative evaluated at this point to the variance through equation (\ref{eq:equal2pt}). The fRG can also be used to compute the exponential decay rate of the covariance in equilibrium through equations (\ref{eq:lambdadef}) \& (\ref{eq:2-pointfunc}). We concluded this section with a computation of the connected 3- and 4-point function in equilibrium.

In part two of this series we will explicitly solve the equations derived here and demonstrate that the fRG offers a practical, computationally faster alternative to established techniques for describing dynamics in and out of equilibrium. We will also explore the validity of the physical relevance of the fRG derivative expansion as temperature is lowered suggesting that it is only applicable in the moderate to high temperature regime.

\section*{Acknowledgements}
AW is funded by the EPSRC under Project 2120421. GR would like to thank Nikos Tetradis for very useful discussions at the early stages of this project, Julien Serreau for providing much insight on fRG computations and Gabriel Moreau for sharing his PhD thesis, containing many new results on the application of the fRG to the Langevin equation.   

\appendix

\section{\label{app:1D=2D}1-D Langevin equation as radial separation of two particles in 2-D or 3-D}
Consider two equal mass particles moving in the same thermal bath with an even\footnote{This precludes our polynomial and unequal LJ potentials} interaction potential between them. The Langevin equations now look like in vector notation:
\begin{eqnarray}
\dot{\vec{x}}_{1} + \varepsilon\grad_{\vec{x}_{1}} V(\norm{\vec{x}_1 - \vec{x}_2}) &=& \vec{\eta}_{1}(t) \label{eq:x1langevin3D}\\
\dot{\vec{x}}_{2} + \varepsilon\grad_{\vec{x}_{2}} V(\norm{\vec{x}_2 - \vec{x}_1}) &=& \vec{\eta}_{2}(t) \label{eq:x2langevin3D} \\
\left\langle \vec{\eta}_i(t)\vec{\eta}_j(t') \right\rangle &=& 2D \delta_{ij}\delta (t-t') \label{eq:noise1}
\end{eqnarray}
Where $\vec{x}_{i}$ is a 3D position vector:
\begin{equation}
\vec{x}_{i} = \left( x_i,y_i,z_i\right)
\end{equation}
and the choice (\ref{eq:noise1}) is made such that $\eta_{i}(t)$ is white noise with $D = k_{b}T/\gamma$ matching the equilibrium Boltzmann distribution. We can now identify our centre of mass vector $\vec{X}$:
\begin{equation}
\vec{X} \equiv \left(\dfrac{x_{1} + x_{2}}{2},\dfrac{y_{1} + y_{2}}{2},\dfrac{z_{1} + z_{2}}{2} \right) = (X,Y,Z)
\end{equation}
And spherical polar coordinates in terms of the relative seperation of the two particles:
\begin{eqnarray}
x_1 - x_2 &=& r\cdot \cos\theta\sin\phi \\
y_1 - y_2 &=& r\cdot \sin\theta\sin\phi \\
z_1 - z_2 &=& r\cdot \cos\phi \\
r &=& \sqrt{(x_1 - x_2)^2+(y_1 - y_2)^2+(z_1 - z_2)^2} \\
\theta &=& \arctan \left(\dfrac{y_1 - y_2}{x_1 - x_2}\right),\quad 0 \leq \theta < 2\pi\\
\phi &=& \arccos \left(\dfrac{z_1 - z_2}{r}\right),\quad 0\leq \phi \leq \pi
\end{eqnarray}
We can then rewrite equations (\ref{eq:x1langevin3D}) \& (\ref{eq:x2langevin3D}):
\begin{eqnarray}
\dot{r} + \varepsilon_r\dfrac{\partial}{\partial r}V(r) &=& \eta_r(t) \label{eq:rlangevin3D}\\
\dot{\theta}\cdot r\sin\phi  &=& \eta_\theta(t)\label{eq:thetalangevin3D}\\
\dot{\phi}\cdot r  &=& \eta_\phi(t)\label{eq:philangevin3D}\\
\left\langle \vec{\eta}_i(t)\vec{\eta}_j(t') \right\rangle &=& 2D_{r} \delta_{ij}\delta (t-t') \\
\dot{\vec{X}} &=& \xi_{\vec{X}}(t) \\
\left\langle \vec{\xi}_i(t)\vec{\xi}_j(t') \right\rangle &=& 2D_{\vec{X}} \delta_{ij}\delta (t-t')
\end{eqnarray}
Where we have introduced effective parameters related to the original ones:
\begin{eqnarray}
\varepsilon_r &\equiv & m/\gamma_r , \gamma_r \equiv \gamma /2 \Rightarrow D_{r} \equiv  2\cdot D_{phys} = \dfrac{k_bT}{\gamma_r}\nonumber \\
\\
D_{X} &\equiv & \dfrac{1}{2}\cdot D_{phys} = \dfrac{k_bT_X}{\gamma} \Leftrightarrow  T_X \equiv  \dfrac{1}{2}\cdot T_{phys}
\end{eqnarray} 
Unfortunately equations (\ref{eq:thetalangevin3D}) \& (\ref{eq:philangevin3D}) depend on r and do not nicely decouple however equation (\ref{eq:rlangevin3D}) is identical to (\ref{eq:langevin}). This means that the 1-D motion of a single particle in a global potential is equivalent to the change in radial seperation between two equal mass particles with an equivalent interaction potential. The centre of the mass of the two particles is described by a simple diffusion equation i.e. the centre of mass goes on a random walk in the dimension of the problem.

\section{\label{app:equil-flow} The equilibrium flow equation} 

In equilibrium, all equal-time expectation values can be generated by the generating function \\
\begin{equation}\label{equil1}
Z(J)=\int dx  \,  e^{-2 V(x)/\Upsilon + Jx} 
\end{equation}
in a manner directly analogous to that described in the text but with functional derivatives replaced by ordinary derivatives w.r.t. $J$. In a spirit identical to the renormalisation group but in the simpler setting of one degree of freedom, we can define a modified generating functional \cite{Guilleux2015}
\begin{equation}
Z_k(J)=\int dx \, e^{-2 V(x)/\Upsilon - \frac{1}{2}R(k)x^2+Jx}
\end{equation}    
with an additional quadratic term controlled by an arbitrary function $R(k)$ of a parameter $k$, satisfying $\lim\limits_{k \rightarrow 0} R(k) = 0$, giving back the original $Z(J)$. Correlation functions are generated by $W_k(J)=\ln Z_k(J)$ via
\begin{equation}\label{app:correlations}
\chi_k \equiv \langle x\rangle_k=\frac{\partial W_k(J)}{\partial J}\,,\quad \langle x^2\rangle_k - \chi^2_k=\frac{\partial^2 W_k(J)}{\partial J^2}
\end{equation}     
e.t.c. In the limit $k=0$ and after setting $J=0$ the usual predictions of the equilibrium Boltzmann distribution are recovered.   

The source $J$ has been considered as an external, independent variable controlling expectation values such as $\chi$ and higher correlators. One could also consider $\chi$ as the independent variable, solving $\chi = \partial W/\partial J$ for $J(\chi)$ and defining the effective potential $U(\chi)$ via a Legendre transform 
\begin{equation}
\Gamma_k(\chi) + W_k(J) = J\chi -\frac{1}{2}R(k)\chi^2
\end{equation}  
with
\begin{equation}
\Gamma(\chi) \equiv 2 U(\chi)/\Upsilon 
\end{equation}            
Note that 
\begin{equation}\label{Legendre}
\frac{\partial \Gamma_k}{\partial \chi } = J_k - R(k)\chi
\end{equation}
implying that the minimum of the effective potential defines the equilibrium expectation value of $x$ (at $J=0$ and $k=0$).

The dependence of the generating function $W_k(J)$ on $k$ can be easily obtained as   
\begin{equation}
\partial_k W_k(J)=-\frac{1}{2}\partial_kR\left[\frac{\partial^2 W_k(J)}{\partial J^2}+\left(\frac{\partial W_k(J)}{\partial J}\right)^2\right]
\end{equation}
which is an ``RG equation'' for $W_k(J)$. We can also obtain an an equation determining how $\Gamma_k(\chi)$ runs with k. Reciprocally, taking $\chi$ as the independent variable, $J$ becomes a function of $\chi$ and $k$. Taking a $k$ derivative of (\ref{Legendre}) at fixed $\chi$ we obtain 
\begin{equation}
\partial_k \Gamma_k(\chi) = \frac{1}{2}\partial_k R \frac{\partial^2 W_k}{\partial J^2} 
\end{equation}            
To express the rhs in terms of $\Gamma_k(\chi)$, consider the first relation of (\ref{app:correlations}). Taking a $\chi$ derivative we find 
\begin{equation}
\left(\frac{\partial^2\Gamma_k}{\partial\chi^2}+R\right)\frac{\partial^2 W_k}{\partial J^2} =1
\end{equation}   
Hence, the ``RG flow'' of $\Gamma$ is determined by 
\begin{equation}\label{app:Gamma-flow}
\partial_k\Gamma_k(\chi)=\frac{1}{2}\partial_kR\left(\frac{\partial^2\Gamma}{\partial\chi^2}+R \right)^{-1}
\end{equation} 
Note also that, at $k \rightarrow 0$ 
\begin{equation}
\langle x^2 \rangle - \chi^2=\frac{\Upsilon}{2 \, \partial_\chi^2 U(\chi_{eq})} 
\end{equation}  
and hence the variance at equilibrium is determined by the curvature of the effective potential around its minimum. 

All the above manipulations can be generalized to many or even infinite degrees of freedom and continuum actions, leading to the Wetterich equation (\ref{Wetterich eqn}), which is directly equivalent to (\ref{app:Gamma-flow}), and the relations of section  (\ref{sec:1-point function}). For this work it is important to note that the equilibrium effective potential $U(\chi)$ discussed here obeys the LPA flow equation \emph{exactly} if we choose $R(k)=k$.         

\section{\label{app:2ptfuncderiv}Derivation of the two point function}
We start from equation (\ref{eq:2pointfuncdef}) repeated here for clarity:
\begin{eqnarray}
\left(\dfrac{d^2}{dt^2} -  Q(t) \right)G(t,t') &=& -\dfrac{\Upsilon}{P(t)}\delta(t-t')  \label{eq:appendix2pointfuncdef} 
\end{eqnarray}
Where $Q(t) = U(\chi(t))$ is given by (\ref{eq:U = }) and $P(t) = 1$ or $\zeta_{\chi}^2(\chi(t))$ for LPA and WFR respectively. If we now consider the homogeneous version of (\ref{eq:appendix2pointfuncdef}):
\begin{equation}
\ddot{f}(t) -  Q(t)f(t) = 0 \label{eq:appendhomo2ptfunc}
\end{equation}
which generically will have two independent solutions $Y_1(t)$ and $Y_2(t)$ which we would like to obtain. In order to do this we consider what these solutions asymptote to at late times. We know for large t (denoted by T) the system will reach equilibrium (or at least will be asymptotically close to it) for which (\ref{eq:appendhomo2ptfunc}) becomes:
\begin{equation}
\ddot{f}(T) - \lambda^2 f(T) = 0 \label{eq:appendhomo2ptfuncequi}
\end{equation}
as $Q(t)$ asymptotes to $\lambda^2$ as the system approaches equilibrium. $\lambda^2$ is defined as in (\ref{eq:lambdadef}):
\begin{eqnarray}\label{eq:lambda}
\lambda^2 \equiv  
\begin{cases}
V_{,\chi\chi}^{2}\vert, & \text{for LPA}\\[10pt]
\dfrac{V_{,\chi\chi}^{2}\vert}{\zeta_{,\chi}^{4}\vert}, & \text{for WFR}
\end{cases} 
\end{eqnarray}
The notation $\vert$ means we have evaluated the function at $k = 0$ and at equilibrium $\chi = \chi_{eq}$. 
Equation (\ref{eq:appendhomo2ptfuncequi}) has two solutions, one growing and one decaying:
\begin{eqnarray}
Y_1(T) &=& A\exp (\lambda T) \label{eq:appendY1equi}\\
Y_2(T) &=& B\exp (-\lambda T)\label{eq:appendY2equi}
\end{eqnarray}
We can now consider the Wronskian $\mathcal{\mathcal{W}}$ which in our case must be constant for all time:
\begin{eqnarray}
\mathcal{W}(t) &\equiv & Y_1(t)\dot{Y}_2(t) - \dot{Y}_1(t)Y_2(t) = \text{constant} \\
\mathcal{W}(T) &=& -2AB\lambda \\
\Rightarrow \mathcal{W}(t) &=& -2AB\lambda \label{eq:appendWronskianconst}
\end{eqnarray}
We will make use of this fact later. \\
Substituting the ansatz $G(t,t') = Y_1(t)F(t,t')$ where F is some function to be determined into (\ref{eq:appendix2pointfuncdef}) we obtain:
\begin{eqnarray}
\dot{F}(t,t') = \dfrac{1}{Y_{1}^{2}(t)}\left[-\Upsilon\dfrac{Y_1(t')}{P(t')}\theta (t-t') + C_1(t')\right] \nonumber \\
\label{eq:append Fdot}
\end{eqnarray}
where $\theta (t-t')$ is the Heaviside step function and $C_1(t')$ is a `constant' of integration function to be determined. If we now integrate (\ref{eq:append Fdot}) we obtain the following expression for $G(t,t')$:
\begin{eqnarray}
G(t,t') &=& -\Upsilon\dfrac{Y_1(t)}{P(t')}\left[\theta (t-t') \int_{t'}^{t} \dfrac{Y_1(t')}{Y_{1}^{2}(u)} du + C_2(t')\right] \nonumber \\
& & \quad  +~C_1(t')Y_1(t)\int^{t} \dfrac{du}{Y_{1}^{2}(u)} \label{eq:appendGwithints}
\end{eqnarray}
where $C_2(t')$ is another `constant' of integration function to be determined. To compute the integrals in (\ref{eq:appendGwithints}) we note that by the definition of the Wronskian:
\begin{eqnarray}
Y_1(t)\int^t \dfrac{\mathcal{W}(u)}{Y_{1}^{2}(u)}du = \mu Y_1(t) +  Y_2(t)
\end{eqnarray}
where $\mu$ is simply a constant of integration. As the Wronskian is constant we simply write:
\begin{eqnarray}
Y_1(t)\int^t \dfrac{du}{Y_{1}^{2}(u)} = \dfrac{1}{-2AB\lambda}\left[\mu Y_1(t) + Y_2(t)\right]
\end{eqnarray}
Such that (\ref{eq:appendGwithints}) becomes:
\begin{eqnarray}
G(t,t') &=& \dfrac{\Upsilon}{2AB\lambda P(t')}\Big\lbrace\bar{C}_1(t')Y_2(t) + \bar{C}_2(t')Y_1(t)  \nonumber \\
 & & \quad +~\theta (t-t')\left[Y_1(t')Y_2(t) - Y_1(t)Y_2(t')\right]\Big\rbrace \nonumber \\ \label{eq:append GCbar}
\end{eqnarray}
where $C_1$ and $C_2$ have been rescaled to $\bar{C}_1$ and $\bar{C}_2$ in order to absorb some irrelevant constant factors. We note that the functions $\bar{C}_i$ can only be linear combinations of $Y_1$ and $Y_2$:
\begin{eqnarray}
\bar{C}_1(t') &\equiv &  \alpha ~Y_1(t') + \beta ~Y_2 (t') \\
\bar{C}_2(t') &\equiv &  \gamma ~Y_1(t') + \delta ~Y_2 (t') 
\end{eqnarray}
where the constants $\alpha$, $\beta$, $\gamma$ and $\delta$ will be determined later\footnote{N.B. the $\delta$ here should not to be confused with the dirac delta function}. Combining all this together we obtain the most general solution:
\begin{eqnarray}
G(t,t') &=& \dfrac{\Upsilon}{2\lambda P(t')}\dfrac{1}{AB}\Big\lbrace \left[\alpha + \theta (t-t') \right] Y_1(t')Y_2(t) \nonumber \\
&& \quad \quad \quad +~\beta ~Y_2(t')Y_2(t)  + \gamma ~Y_1(t')Y_1(t)\nonumber \\
 & & \quad  \quad  \quad+~\left[\delta - \theta (t-t')\right]Y_2(t')Y_1(t)\Big\rbrace \nonumber \\ \label{eq:append G general}
\end{eqnarray}
To obtain the values of the constants we must impose physical conditions:
\begin{enumerate}
\item The variance $G(t,t)$ should remain finite as $t\rightarrow \infty$ i.e. an equilibrium distribution exists at late times \\
$\Rightarrow \gamma = 0$
\item Variance $G(t,t)$ should approach the correct equilibrium distribution $G_{eq}$ at late times T \\
$\Rightarrow \alpha = 0$
\item Covariance $G(t,0)$ should remain finite as $t\rightarrow \infty$  \\
$\Rightarrow \delta = 1$
\item The initial condition is $G(0,0) \equiv G_{00}$ \\
$\Rightarrow \dfrac{\beta\Upsilon}{2AB\lambda} = \dfrac{P(0)}{Y_2(0)Y_2(0)}\left[G_{00} - \dfrac{Y_1(0)Y_2(0)}{AB}\dfrac{\Upsilon}{2\lambda P(0)}\right]$
\end{enumerate}
Which gives us the two point function:
\begin{eqnarray}
G(t,t') &=& \dfrac{\Upsilon}{2\lambda P(t')}\left[ \theta (t-t')\tilde{Y}_1(t')\tilde{Y}_2(t) + \theta (t'-t)\tilde{Y}_2(t')\tilde{Y}_1(t)\right] \nonumber \\
&& \quad \quad +~ \dfrac{P(0)}{P(t')}\left[G_{00} - \dfrac{\Upsilon}{2\lambda P(0)}\right]\tilde{Y}_2(t')\tilde{Y}_2(t) \label{eq:append finalG}
\end{eqnarray}
where $\tilde{Y}_i(t) \equiv Y_i(t)/Y_i(0)$ are the `normalised' solutions to the homogeneous equation (\ref{eq:appendhomo2ptfunc}). We have also set $A = Y_1(0)$ and $B = Y_2(0)$ which we are free to do.
Equation (\ref{eq:append finalG}) has two important limits: \\
The \textit{Variance} $t'\rightarrow t$:
\begin{eqnarray}
\textbf{Var}(x) \equiv G(t,t) &=& \dfrac{\Upsilon}{2\lambda P(t)} \tilde{Y}_1(t)\tilde{Y}_2(t)  \nonumber \\
&& +~ \dfrac{P(0)}{P(t)}\left[G_{00} - \dfrac{\Upsilon}{2\lambda P(0)}\right]\tilde{Y}_{2}^{2}(t)\nonumber \\
 \label{eq:append Variance}
\end{eqnarray}
and the \textit{Covariance} $t'\rightarrow 0$, $t > 0$:
\begin{eqnarray}
\textbf{Cov}(x(0)x(t)) \equiv G(t,0) = G_{00}\tilde{Y}_2(t) \label{eq:append Covariance}
\end{eqnarray}
Equations (\ref{eq:append Variance}) \& (\ref{eq:append Covariance}) are the main results of this appendix. 
\bibliography{library}
\end{document}